\documentclass[aps,prd,twocolumn,superscriptaddress,nofootinbib,showkeys]{revtex4}
\usepackage{amssymb,amsmath,graphicx,enumerate}
\usepackage{subfigure}
\usepackage{dcolumn,booktabs}
\usepackage{longtable}
\usepackage[dvipdfm,colorlinks=true,citecolor=blue,pdfstartview=FitH]{hyperref}

\def\bea{\begin{equation}}
\def\eea{\end{equation}}
\newcommand{\rt}{Regge trajectory}
\newcommand{\rts}{Regge trajectories}
\newcommand{\bfr}{{\bf r}}

\newcommand{\bfpa}{{|\bf p|}}
\newcommand{\gev}{{\rm GeV}}

\newcommand{\sse}{spinless Salpeter equation}

\newcommand{\cltb}{$\bar{3}_c$}
\newcommand{\cltba}{\bar{3}_c}
\newcommand{\cls}{$6_c$}
\newcommand{\dqs}{$(qq')$}

\begin{document}
\title{Regge trajectories for the heavy-light diquarks}
\author{Jiao-Kai Chen}
\email{chenjk@sxnu.edu.cn, chenjkphy@outlook.com}
\affiliation{School of Physics and Information Engineering, Shanxi Normal University, Taiyuan 030031, China}
\author{Xia Feng}
\email{sxsdwxfx@163.com}
\affiliation{School of Physics and Information Engineering, Shanxi Normal University, Taiyuan 030031, China}
\author{Jia-Qi Xie}
\email{1462718751@qq.com}
\affiliation{School of Physics and Information Engineering, Shanxi Normal University, Taiyuan 030031, China}
\date{\today}

\begin{abstract}
We attempt to apply the Regge trajectory approach to the heavy-light diquarks composed of one heavy quark and one light quark. However, we find that the direct application of the usual Regge trajectory formula for the heavy-light mesons and baryons fails. In order to correctly estimate the masses of the heavy-light diquarks, it is needed to consider the light quark mass correction and the parameter $C$ in the Cornell potential within the Regge trajectory formula. By using the modified Regge trajectory formulas, we are able to estimate the masses of the heavy-light diquarks $(cu)$, $(cs)$, $(bu)$ and $(bs)$, which agree with other theoretical results.
It is illustrated that the heavy-light diquarks satisfy the universal descriptions irrespective of heavy quark flavors, similar to other heavy-light systems such as the heavy-light mesons, the heavy-light baryons composed of one heavy quark (diquark) and one light diquark (quark), and the heavy-light tetraquarks composed of one heavy diquark (antidiquark) and one light antidiquark (diquark).
The diquark Regge trajectory provides a new and very simple approach for estimating the spectra of the heavy-light diquarks.
\end{abstract}

\maketitle

%%%%%%%%%%%%%%%%%%%%%%%%%%%%%%%%%%%%%%%%%%%%%%%%%%%%%%%%%%%%%%%%%%%%%%%%%%

\section{Introduction}
Phenomenology suggests that diquark correlations might play a material role in the formation of exotic tetraquarks and pentaquarks \cite{Jaffe:2004ph,Liu:2019zoy,Esposito:2016noz,Olsen:2017bmm,Guo:2019twa}.
Diquark substructure affects the static properties of baryons, tetraquarks and pentaquarks \cite{Gell-Mann:1964ewy,Lichtenberg:1967zz,
Anselmino:1992vg,Jaffe:2004ph,Barabanov:2020jvn,Selem:2006nd,Wilczek:2004im,Jaffe:2003sg}.
In the diquark picture, diquarks are constituents of baryons \cite{Lichtenberg:1967zz,Lu:2017meb,Ida:1966ev}, tetraquarks \cite{Liu:2019zoy,Jaffe:1976ig,Esposito:2016noz,Maiani:2004vq} and pentaquarks \cite{Liu:2019zoy,Jaffe:2004ph,Esposito:2016noz,Lebed:2015tna}.

The spectra of the heavy-light diquarks composed of one heavy quark and one light quark have been studied using various approaches. In Ref. \cite{Faustov:2021hjs}, the heavy-light diquark masses are calculated by using the Sch\"{o}dinger-type quasipotential equations. In Refs. \cite{Yin:2021uom,Gutierrez-Guerrero:2021rsx,
Gutierrez-Guerrero:2019uwa,Yu:2006ty}, the masses of different kinds of diquarks are calculated by applying the Bethe-Salpeter equation. In Refs. \cite{Giannuzzi:2019esi,Ferretti:2019zyh}, the mass spectra of the diquarks are obtained in the potential model. In Ref. \cite{Gutierrez-Guerrero:2021fuj}, the masses of different types of diquarks are calculated within a non-relativistic potential model. In Ref. \cite{Wang:2010sh,Kleiv:2013dta,deOliveira:2023hma}, the heavy-light diquark masses are obtained from QCD sum rules.

The {\rt} is one of the effective approaches for studying hadron spectra \cite{Regge:1959mz,Chew:1962eu,Nambu:1978bd,Ademollo:1969nx,Baker:2002km,
Brodsky:2006uq,Forkel:2007cm,Filipponi:1997vf,brau:04bs,Brisudova:1999ut,
Guo:2008he,Feng:2023ynf,Lovelace:1969se,Irving:1977ea,Collins:1971ff,
Inopin:1999nf,MartinContreras:2020cyg,Sonnenschein:2018fph,Afonin:2023lfi,
MartinContreras:2022ebv,Patel:2022hhl,Abreu:2020ttf,Badalian:2019lyz,
FolcoCapossoli:2019imm,Chen:2023web,Chen:2023djq,Chen:2022flh,Afonin:2014nya,Jakhad:2023mni,
Chen:2014nyo,Chen:2017fcs,Veseli:1996gy,Jia:2018vwl,Anisovich:2000kxa,
Ebert:2009ub,Masjuan:2012gc,Badalian:2019dny,Chen:2018hnx,Chen:2018bbr,Chen:2018nnr,Chen:2021kfw}.
Although diquarks are colored states and not physical \cite{Jaffe:2004ph}, we have attempted to apply the {\rt} approach\footnote{The {\rts} are commenly plotted in the $(M^2,\,x)$ plane or in the $(x,\,M^2)$ plane, where $x=l,\,n_r$. For simplicity, the figures plotted in the $(M,\,x)$ plane, the $(M-m_R,\,x)$ plane and the $((M-m_R)^2,\,x)$ plane are also called the {\rts}.} to discuss the doubly heavy diquarks in Ref. \cite{Feng:2023txx}. The obtained results agree with the theoretical predictions calculated by other methods.
The diquark Regge trajectory offers a new and simple approach for estimating the spectra of diquarks.
Furthermore, it is expected that the diquark Regge trajectory can provide an easy method for investigating the $\rho-$mode excitations of baryons, tetraquarks and pentaquarks containing diquarks \cite{chen:2023rho}.
In this study, we attempt to apply the {\rt} approach to investigate the heavy-light diquarks. We find that the direct use of the {\rt} formula for the heavy-light mesons and baryons fails, and the {\rt} formula needs to be modified to fit the {\rts} for the heavy-light diquarks.

The paper is organized as follows: In Sec. \ref{sec:rgr}, the {\rt} relations are obtained from the spinless Salpeter equation (SSE). In Sec. \ref{sec:rtdiquark}, we investigate the {\rts} for the heavy-light diquarks. The conclusions are presented in Sec. \ref{sec:conc}.

\section{{\rt} relations for the heavy-light diquarks}\label{sec:rgr}

In this section, we attempt to seek unified {\rt} formulas for the heavy-light mesons $q\bar{q}'$ ($q=c,b$, $q'=u,d,s$) and the heavy-light diquarks $(qq')$ [where $\bar{q}'$ is the antiquark of $q'$].

\subsection{SSE}
The {\sse} (SSE) \cite{Godfrey:1985xj,Ferretti:2019zyh,Durand:1981my,Durand:1983bg,Lichtenberg:1982jp,Jacobs:1986gv} reads as
\begin{eqnarray}\label{qsse}
M\Psi_{d,m}({\bfr})=\left(\omega_1+\omega_2\right)\Psi_{d,m}({\bfr})+V_{d,m}\Psi_{d,m}({\bfr}),
\end{eqnarray}
where $M$ is the bound state mass (diquark or meson). $\Psi_{d,m}({\bfr})$ are the diquark wave function and the meson wave function, respectively. $V_{d,m}$ are the diquark potential and the meson potential, respectively, see Eq. (\ref{potential}). $\omega_1$ is the relativistic energy of quark $q$, and $\omega_2$ is of quark $q'$ or antiquark $\bar{q}'$,
\bea\label{omega}
\omega_i=\sqrt{m_i^2+{\bf p}^2}=\sqrt{m_i^2-\Delta}\;\; (i=1,2).
\eea
$m_1$ and $m_2$ are the effective masses of heavy quark $q$ and light quark $q'$ (or antiquark $\bar{q}'$), respectively.

According to the $SU_c(3)$ color symmetry, a meson is a color singlet composed of one quark in $3_c$ and one antiquark in {\cltb}. The diquark composed of two quarks in $3_c$ is a color antitriplet {\cltb} or a color sextet $6_c$.
Only the {\cltb} representation of $SU_c(3)$ is considered in the present work and the {\cls} representation \cite{Weng:2021hje,Praszalowicz:2022sqx} is not considered. %%
In $SU_c(3)$, there is attraction between quark pairs $(qq')$ in the color antitriplet channel and this is just twice weaker than in the color singlet $q\bar{q}'$ in the one-gluon exchange approximation \cite{Esposito:2016noz}.
It is introducing a factor $1/2$. One would expect $1/2$ to be a global factor since it comes from the color structure of the wave function and it is common to extend this factor to the whole potential describing the quark-quark interaction \cite{Debastiani:2017msn}.
Following Ref. \cite{Ferretti:2019zyh}, we employ the potential
\bea\label{potential}
V_{d,m}=-\frac{3}{4}\left[-\frac{4}{3}\frac{\alpha_s}{r}+{\sigma}r+C\right]
\left({\bf{F}_i}\cdot{\bf{F}_j}\right)_{d,m},
\eea
where $\alpha_s$ is the strong coupling constant of the color Coulomb potential. $\sigma$ is the string tension. $C$ is a fundamental parameter \cite{Gromes:1981cb,Lucha:1991vn}. The part in the bracket is the Cornell potential \cite{Eichten:1974af}. ${\bf{F}_i}\cdot{\bf{F}_j}$ is the color-Casimir,
\bea\label{mrcc}
\langle{(\bf{F}_i}\cdot{\bf{F}_j})_d\rangle=-\frac{2}{3},\quad
\langle{(\bf{F}_i}\cdot{\bf{F}_j})_m\rangle=-\frac{4}{3}.
\eea
The value of $({\bf{F}_i}\cdot{\bf{F}_j})_{d}$ is half of $({\bf{F}_i}\cdot{\bf{F}_j})_{m}$, which agrees with the relation
\cite{Faustov:2021hjs,Godfrey:1985xj,Debastiani:2017msn}
\bea\label{halfr}
V_{d}=\frac{V_{m}}{2}.
\eea
According to Eqs. (\ref{qsse}), (\ref{potential}), and (\ref{halfr}), we see that the diquark and meson are described in an unified approach. Therefore, it is expected that the heavy-light diquarks and the heavy-light mesons can be described universally by the {\rt} approach.

\subsection{{\rt} relations}
In case of the heavy-light mesons, there is $m_{1}\to\infty$. Using the limit $m_1\to\infty$ and Eq. (\ref{qsse}), we have
\bea\label{expem}
M=m_1+\langle\omega_2\rangle+\langle{V_{d,m}}\rangle.
\eea
The expectation values are understood to be taken with respect to the normalized eigenstates of Eq. (\ref{qsse}). For large $r$, we neglect the color Coulomb potential. Using the relativistic virial theorem \cite{Lucha:1989jf}, we have from Eq. (\ref{expem})
\bea\label{experd}
M=m_1-\frac{3}{4}C\langle{(\bf{F}_i}\cdot{\bf{F}_j})_{d,m}\rangle
+\left\langle\sqrt{m_2^2+{\bf{p}}^2}
+\frac{{\bf{p}}^2}{\sqrt{m_2^2+{\bf{p}}^2}}\right\rangle.
\eea
Using $\left\langle\sqrt{m_2^2+{\bf{p}}^2}\right\rangle
\le\sqrt{m_2^2+\langle{{\bf{p}}^2}\rangle}$, we have the approximate relation,
\bea\label{experf}
M{\approx}m_1-\frac{3}{4}C\langle{(\bf{F}_i}\cdot{\bf{F}_j})_{d,m}\rangle
+\sqrt{m_2^2+\langle{{\bf{p}}^2}\rangle}
+\frac{\langle{{\bf{p}}^2}\rangle}{\sqrt{m_2^2+\langle{{\bf{p}}^2}\rangle}}.
\eea
Using Eq. (\ref{experf}) and the parameters in Eq. (\ref{param}), we can estimate the order of momenta of light quark in the heavy-light mesons. For the $1^1S_0$ state $D^{\pm}$, $\langle{\bfpa}\rangle{\sim}0.28$ {\gev}. For the $2^1S_0$ state $D_0(2500)^0$, $\langle{\bfpa}\rangle{\sim}0.65$ {\gev}. For the $1^1S_0$ state $B^{\pm}$, $\langle{\bfpa}\rangle{\sim}0.33$ {\gev}. For the $1^3P_2$ state $B^{\ast}_2(5747)$, $\langle{\bfpa}\rangle{\sim}0.57$ {\gev}. The momenta of light quark increases with $l$ or $n_r$, where $l$ ($l=0,1,2,\cdots$) is the orbital angular momentum and $n_r$ ($n_r=0,1,2,\cdots$) is the radial quantum number. Comparing with the momenta of light quarks, it is reasonable that the mass of the light quark is regarded as being small for the orbitally and radially excited states of the heavy-light mesons.

The mass of the light antiquark is assumed to approach zero, $m_2\to0$ in Refs. \cite{Veseli:1996gy,Chen:2017fcs} or is taken as being very small and is considered by correction term in Refs. \cite{Selem:2006nd,Chen:2014nyo,Afonin:2014nya,Sonnenschein:2018fph,Jakhad:2023mni}. In the limit $m_1\to\infty$ and $m_2\to0$, Eq. (\ref{qsse}) is reduced to be
\begin{eqnarray}\label{qssenr}
M\Psi_m({\bfr})=\left[m_1+{\bfpa}+V_m\right]\Psi_m({\bfr}).
\end{eqnarray}
By employing the Bohr-Sommerfeld quantization approach \cite{brau:04bs,brsom}, we have from Eq. (\ref{qssenr})
\bea\label{rtfsm}
M{\sim}2\sqrt{\sigma}\sqrt{l},\;M{\sim}\sqrt{2{\pi}\sigma}\sqrt{n_r}.
\eea
Using Eq. (\ref{rtfsm}), the parameterized formula can be written as \cite{Chen:2023web,Afonin:2014nya,Jakhad:2023mni,Chen:2014nyo,
Chen:2017fcs,Veseli:1996gy,Jia:2018vwl}
\bea\label{rtmeson}
M=m_R+\beta_x\sqrt{x+c_{0x}},\,(x=l,\,n_r).
\eea
The parameter in Eq. (\ref{rtmeson}) reads as \cite{Chen:2023web,Chen:2021kfw}
\bea\label{massform}
\beta_x=c_{fx}c_xc_{d,m}.
\eea
The constants $c_{x}$ and $c_{m}$ are
\bea\label{cxcons}
c_{m}=\sqrt{\sigma},\quad c_l=2,\quad c_{n_r}=\sqrt{2\pi}.
\eea
$c_{d}$ is in Eq. (\ref{sigma}). Both $c_{fl}$ and $c_{fn_r}$ are theoretically equal to one and are fitted in practice. For the heavy-light mesons, the common choice of $m_R$ is \cite{Selem:2006nd,Chen:2023web,Jakhad:2023mni,Chen:2014nyo,
Chen:2017fcs,Veseli:1996gy,Jia:2018vwl}
\bea\label{mrm1}
m_R=m_1.
\eea
%%%
In Eqs. (\ref{massform}) and (\ref{cxcons}), $m_R$, $c_x$, $c_{fx}$ and $\sigma$ are universal for the heavy-light mesons. $c_{0x}$, which varies with different {\rts}, is determined by fitting the given {\rt}.

Similar to the preceding discussions, the {\rt} formula for the heavy-light diquarks can be obtained, which has the same form as that for the heavy-light mesons, i.e., Eq. (\ref{rtmeson}) with (\ref{massform}) and (\ref{cxcons}) and the $\sigma$ in (\ref{cxcons}) should be divided by a factor of 2,
\bea\label{sigma}
c_{d}=\sqrt{\frac{\sigma}{2}} \text{ for diquark}.
\eea

By fitting both the heavy-light meson and the heavy-light diquark {\rts}, we find that the usual {\rt} Eq. (\ref{rtmeson}) with (\ref{mrm1}), which is obtained in the limit $m_1\to\infty$ and $m_2\to0$, cannot give agreeable results of diquarks. In reality, even the light quarks are massive. $C$ should be considered according to (\ref{potential}) and (\ref{qssenr}). Moreover, we find that considering only one of the light quark mass correction and the parameter $C$ cannot give acceptable results. Therefore, both the light quark mass correction and the parameter $C$ are needed to be considered.
There are different ways to include the light quark mass correction \cite{Afonin:2014nya,Selem:2006nd,Chen:2014nyo,Sonnenschein:2018fph,
MartinContreras:2020cyg}. According to our knowledge, the light quark mass correction is not obtained from Eq. (\ref{qsse}) due to its complexity.
In Ref. \cite{Afonin:2014nya}, the authors propose $M=m_1+m_2+\sqrt{a(n_r+{\alpha}l+b)}$ based on the string model, which differs from Eq. (\ref{rtmeson}) mainly in the term $m_2$.
According to the discussions in Refs. \cite{Afonin:2014nya,Chen:2022flh}, by including the parameter $C$, we have Eq. (\ref{rtmeson}) with
\bea\label{rtft}
m_R=m_1+m_2+C_{d,m},\;C_d=\frac{C}{2},\; C_m=C.
\eea
In Ref. \cite{Selem:2006nd}, $M=m_1+\sqrt{{\sigma'}l/2}+2^{1/4}{\kappa}l^{-1/4}m_2^{3/2}$, which differs from (\ref{rtmeson}) mainly in the corrections of $m_2$, is obtained by the computer simulations for the heavy-light systems based on the loaded flux tube model.
According to the results in \cite{Selem:2006nd,Sonnenschein:2018fph}, by simple algebra calculations and including the parameter $C$, we have
\bea\label{mrtf}
M=m_R+\sqrt{\beta_x^2(x+c_{0x})+\kappa_{x}m^{3/2}_2(x+c_{0x})^{1/4}}
\eea
if $m_2{\ll}M$. Eq. (\ref{mrtf}) agrees with the results in Refs. \cite{Selem:2006nd,Sonnenschein:2018fph}. In Eq. (\ref{mrtf}),
\bea\label{mrfp}
m_R=m_1+C_{d,m},\quad \kappa_x=\frac{4}{3}\sqrt{{\pi}\beta_x},
\eea
where $\beta_x$ is in (\ref{massform}), $C_{d,m}$ are in (\ref{rtft}). As $m_2=0$, these two modified formulas, formulas (\ref{rtmeson}) with (\ref{rtft}) and (\ref{mrtf}) with (\ref{mrfp}), become identical. As $m_2=0$ and $C$ is neglected, these two modified formulas reduce to the usual {\rt} formula for the heavy-light mesons, i.e., (\ref{rtmeson}) with (\ref{mrm1}).

The {\rt} relation for the doubly heavy diquarks has the same form as the {\rt} relation for the doubly heavy mesons \cite{Feng:2023txx}. Different from the doubly heavy diquark case, the usual {\rt} formula (\ref{rtmeson}) with (\ref{mrm1}) for the heavy-light mesons cannot be applied directly to the heavy-light diquarks. [Because $m_2+C$ is small for $m_1$ and then $m_R=m_1$ is a good approximation of $m_R=m_1+m_2+C$, the usual {\rt} formula (\ref{rtmeson}) with (\ref{mrm1}) can work well for the heavy-light mesons.]
The unified description of the heavy-light mesons and the heavy-light diquarks demands more rigorous {\rt} relations. By fitting masses of the heavy-light mesons and the heavy-light diquarks, we find that the formula (\ref{rtmeson}) with (\ref{rtft}) and the formula (\ref{mrtf}) with (\ref{mrfp}) are acceptable. They are justified not only by that they are obtained from the string model or from the loaded flux tube model but also by that they can produce the heavy-light mesons masses and the heavy-light diquark masses which are in agreement with other theoretical predictions, see details in Sec. \ref{sec:rtdiquark}. Different from other formulas, formulas (\ref{rtmeson}) with (\ref{rtft}) and (\ref{mrtf}) with (\ref{mrfp}) include not only the quark masses $m_1$ and $m_2$ but also the fundamental parameter $C$ in the Cornell potential. Moreover, the half relation (\ref{halfr}) is explicitly included in these two formulas through $\sigma/2$ and $C/2$. These two {\rt} formulas present unified descriptions of the heavy-light mesons and the heavy-light diquarks.

\section{{\rts} for the heavy-light diquarks}\label{sec:rtdiquark}

In this section, the {\rts} for the heavy-light diquarks $(cu)$, $(cs)$, $(bu)$ and $(bs)$ are investigated.

\subsection{Preliminary}\label{subsec:pre}

\begin{table*}[!phtb]
\caption{The completely antisymmetric states for the diquarks in {\cltb} and in $6_c$ \cite{Feng:2023txx}. $j$ is the spin of the diquark {\dqs}, $s$ denotes the total spin of two quarks, $l$ represents the orbital angular momentum. $n=n_r+1$, $n_r$ is the radial quantum number, $n_r=0,1,2,\cdots$. }  \label{tab:dqstates}
\centering
\begin{tabular*}{0.8\textwidth}{@{\extracolsep{\fill}}ccccc@{}}
\hline\hline
 Spin of diquark & Parity  &  Wave state  &  Configuration    \\
( $j$ )          & $(j^P)$ & $(n^{2s+1}l_j)$  \\
\hline
j=0              & $0^+$   & $n^1s_0$         & $[qq']^{{\cltba}}_{n^1s_0}$,\; $\{qq'\}^{{6_c}}_{n^1s_0}$ \\
                 & $0^-$   & $n^3p_0$         & $[qq']^{{\cltba}}_{n^3p_0}$,\; $\{qq'\}^{{6_c}}_{n^3p_0}$       \\
j=1              & $1^+$   & $n^3s_1$, $n^3d_1$   & $\{qq'\}^{{\cltba}}_{n^3s_1}$,\;    $\{qq'\}^{{\cltba}}_{n^3d_1}$,\;
$[qq']^{6_c}_{n^3s_1}$,\;    $[qq']^{{6_c}}_{n^3d_1}$\\
                 & $1^-$   & $n^1p_1$, $n^3p_1$   &
$\{qq'\}^{{\cltba}}_{n^1p_1}$,\; $[qq']^{{\cltba}}_{n^3p_1}$, \;          $[qq']^{6_c}_{n^1p_1}$,\; $\{qq'\}^{6_c}_{n^3p_1}$ \\
j=2              & $2^+$   & $n^1d_2$, $n^3d_2$         &  $[qq']^{{\cltba}}_{n^1d_2}$,\; $\{qq'\}^{{\cltba}}_{n^3d_2}$,\;
$\{qq'\}^{6_c}_{n^1d_2}$,\; $[qq']^{6_c}_{n^3d_2}$         \\
                 & $2^-$   & $n^3p_2$, $n^3f_2$       &
 $[qq']^{{\cltba}}_{n^3p_2}$,\; $[qq']^{{\cltba}}_{n^3f_2}$,\;
  $\{qq'\}^{6_c}_{n^3p_2}$,\; $\{qq'\}^{6_c}_{n^3f_2}$          \\
$\cdots$         & $\cdots$ & $\cdots$               & $\cdots$  \\
\hline\hline
\end{tabular*}
\end{table*}

The state of diquark $(qq')$ is denoted as $[qq']^{c}_{n^{2s+1}l_j}$ or $\{qq'\}^{c}_{n^{2s+1}l_j}$, see Table \ref{tab:dqstates}. $c={\cltba},\,6_c$. $\{qq'\}$ and $[qq']$ indicate the permutation symmetric and antisymmetric flavor wave functions, respectively. $n=n_r+1$, $n_r=0,1,\cdots$, where $n_r$ is the radial quantum number. $s$ is the total spin of two quarks, $l$ is the orbital quantum number, and $j$ is the spin of the diquark $(qq')$.

Two {\rt} formulas are used to fit the radial and orbital {\rts} for the heavy-light diquarks: (1) Eq. (\ref{rtmeson}) with (\ref{massform}), (\ref{cxcons}), (\ref{sigma}), and (\ref{rtft}) (Fit1); (2) (\ref{mrtf}) with (\ref{mrfp}) (Fit2).
%%%
The quality of a fit is measured by the quantity $\chi^2$ defined by \cite{Sonnenschein:2014jwa}
\bea
\chi^2=\frac{1}{N-1}\sum^{N}_{i=1}\left(\frac{M_{fi}-M_{ei}}{M_{ei}}\right)^2,
\eea
where $N$ is the number of points on the trajectory, $M_{fi}$ is fitted value and $M_{ei}$ is the experimental value or the theoretical value of the $i$-th particle mass. The parameters are determined by minimizing $\chi^2$.
Firstly, using two {\rt} formulas and the masses from PDG \cite{ParticleDataGroup:2022pth} and the masses predicted theoretically \cite{Ebert:2009ua}, we fit the radial and orbital {\rts} for the charmed, charmed-strange, bottom and bottom-strange mesons, respectively. Secondly, we choose the following parameter values \cite{Faustov:2021hjs,Ebert:2002ig} which are used to fit the {\rts} for the heavy-light diquarks,
\begin{align}\label{param}
m_b&=4.88\; {\gev},\quad m_c=1.55\; {\gev},\nonumber\\
m_{u,d}&=0.33\; {\gev},\quad m_s=0.50\; {\gev},\nonumber\\
\sigma&=0.18\; {\gev^2},\quad C=-0.30\; {\gev}.
\end{align}
These parameters are universal for all doubly heavy diquark {\rts} \cite{Feng:2023txx} and for all heavy-light diquark {\rts}. Thirdly, the only free parameter $c_{0x}$ in (\ref{rtmeson}), which varies with different diquark {\rts}, is determined by fitting the corresponding meson {\rts}. For example, the $c_{0n_r}$ in the $[bu]^{{\cltba}}_{1^3p_0}$ {\rt} is calculated by fitting the radial {\rt} for the $1^3P_0$ bottom meson. As all parameters are determined, the diquark masses can be calculated finally.
%%%
There is not compelling reason why $c_{0x}$ obtained by fitting the meson {\rts} can be used directly to calculate the diquark {\rts}. We use this method as a provisional method to determine $c_{0x}$ before finding a better one. It validates this method that the fitted results for the heavy-light diquarks $(cu)$, $(cs)$, $(bu)$ and $(bs)$ agree with the theoretical values obtained by using other approaches, see the discussions in the following subsections.

\begin{table*}[!phtb]
\caption{The fitted values of parameters $c_{fn_r}$, $c_{fl}$, $c_{0n_r}$, and $c_{0l}$. $c_{0n_r}$ ($1^1s_0$) and $c_{0n_r}$ ($1^3s_1$) are obtained by fitting the radial {\rt} for the $1^1s_0$ state and the $1^3s_1$ state, respectively. The $c_{0l}$ ($1^1s_0$), $c_{0l}$ ($1^3s_1$) and $c_{0l}$ ($1^3p_0$) are calculated by fitting the orbital {\rt} for the $1^1s_0$ state, the $1^3s_1$ state and the $1^3p_0$ state, respectively. }  \label{tab:fitparameters}
\centering
\begin{tabular*}{0.9\textwidth}{@{\extracolsep{\fill}}cccccccccc@{}}
\hline\hline
           &   \multicolumn{2}{c}{$(cu)$} &   \multicolumn{2}{c}{$(bu)$} &   \multicolumn{2}{c}{$(cs)$} &   \multicolumn{2}{c}{$(bs)$}   \\
                   & Fit1 &  Fit2  & Fit1   &  Fit2  & Fit1   &  Fit2  & Fit1   &  Fit2 \\
\hline
$c_{fn_r}$           &1.000 & 1.102 & 0.988 & 1.093 &1.016 &1.154 &0.953 &1.086    \\
$c_{fl}$             &1.038 & 1.157 & 0.965 & 1.076 &1.015 &1.162 &0.919 &1.055    \\
%\hline
$c_{0n_r}$ ($1^1s_0$)    &0.065 &0.075 &0.125 &0.145 &0.03 &0.04 &0.08 &0.095    \\
$c_{0n_r}$ ($1^3s_1$)    &0.17  &0.2   &0.155 &0.18  &0.095&0.11 &0.11 &0.125    \\
$c_{0l}$ ($1^1s_0$)      &0.095 &0.11  &0.18 &0.205 &0.055 &0.07 &0.115&0.135    \\
$c_{0l}$ ($1^3s_1$)      &0.19  &0.215 &0.22 &0.255 &0.135 &0.155&0.16 &0.18    \\
$c_{0l}$ ($1^3p_0$)      &0.9   &0.93  &1.18 &1.21  &0.83  &0.855&1.09 &1.11    \\
\hline
\hline
\end{tabular*}
\end{table*}

\subsection{{\rts} for the $(cu)$ and $(bu)$  diquarks}\label{subsec:cbu}

\begin{table}[!phtb]
\caption{The fitted values (in {\gev}) for the diquarks $(cu)$ and $(bu)$ by using the radial {\rts}. $n=n_r+1$, $n_r=0,1,\cdots$. $n_r$ is the radial quantum number. $s$ is the total spin of two quarks, $l$ is the orbital quantum number and $j$ is the spin of diquark. }  \label{tab:bcur}
\centering
\begin{tabular*}{0.47\textwidth}{@{\extracolsep{\fill}}cccccc@{}}
\hline\hline
                      &   \multicolumn{2}{c}{$(cu)$} &   \multicolumn{2}{c}{$(bu)$}   \\
State ($n^{2s+1}l_j$) & Fit1 &  Fit2  & Fit1   &  Fit2 \\
\hline
$1^1s_0$           & 1.92    & 1.91    & 5.32    & 5.32      \\
$2^1s_0$           & 2.51    & 2.47    & 5.85    & 5.82     \\
$3^1s_0$           & 2.81    & 2.78    & 6.14    & 6.12     \\
$4^1s_0$           & 3.05    & 3.03    & 6.37    & 6.36      \\
$5^1s_0$           & 3.25    & 3.24    & 6.57    & 6.57     \\
%%%
$1^3s_1$           & 2.04    & 2.04    & 5.35    & 5.35    \\
$2^3s_1$           & 2.54    & 2.52    & 5.86    & 5.83    \\
$3^3s_1$           & 2.84    & 2.82    & 6.15    & 6.13    \\
$4^3s_1$           & 3.07    & 3.06    & 6.38    & 6.37    \\
$5^3s_1$           & 3.26    & 3.26    & 6.57    & 6.58    \\
\hline
\hline
\end{tabular*}
\end{table}

\begin{table}[!phtb]
\caption{Same as Table \ref{tab:bcur} except by using the orbital {\rts}. }  \label{tab:bcuo}
\centering
\begin{tabular*}{0.47\textwidth}{@{\extracolsep{\fill}}cccccc@{}}
\hline\hline
                      &   \multicolumn{2}{c}{$(cu)$} &   \multicolumn{2}{c}{$(bu)$}   \\
State ($n^{2s+1}l_j$) & Fit1 &  Fit2  & Fit1   &  Fit2 \\
\hline
$1^1s_0$         & 1.92    & 1.92  & 5.31  & 5.30 \\
$1^1p_1$         & 2.38    & 2.36  & 5.69  & 5.67 \\
$1^1d_2$         & 2.63    & 2.61  & 5.92  & 5.90 \\
$1^1f_3$         & 2.83    & 2.81  & 6.09  & 6.08 \\
$1^1g_4$         & 2.99    & 2.98  & 6.24  & 6.24 \\
$1^1h_5$         & 3.14    & 3.14  & 6.38  & 6.38 \\
%%%\hline
$1^3s_1$         & 2.00    & 2.00  & 5.33  & 5.33   \\
$1^3p_2$         & 2.41    & 2.39  & 5.70  & 5.68 \\
$1^3d_3$         & 2.65    & 2.63  & 5.92  & 5.91\\
$1^3f_4$         & 2.84    & 2.83  & 6.10  & 6.09\\
$1^3g_5$         & 3.01    & 3.00  & 6.25  & 6.24\\
$1^3h_6$         & 3.15    & 3.15  & 6.38  & 6.38 \\
\hline\hline
\end{tabular*}
\end{table}

\begin{figure*}[!phtb]
\centering
\subfigure[]{\label{figcca}\includegraphics[scale=0.8]{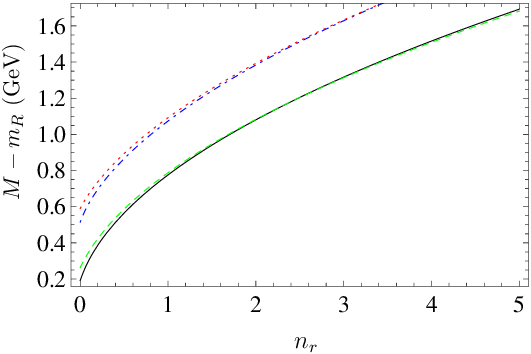}}
\subfigure[]{\label{figcca}\includegraphics[scale=0.8]{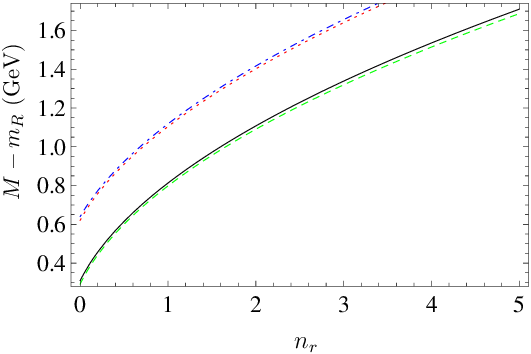}}
\subfigure[]{\label{figcca}\includegraphics[scale=0.8]{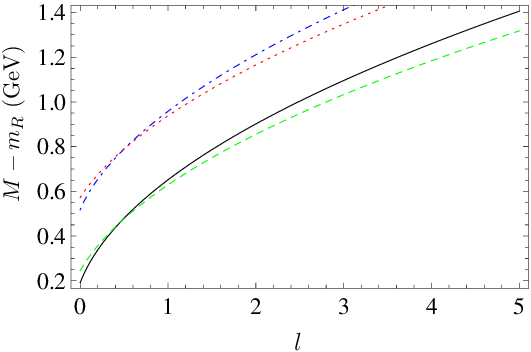}}
\subfigure[]{\label{figcca}\includegraphics[scale=0.8]{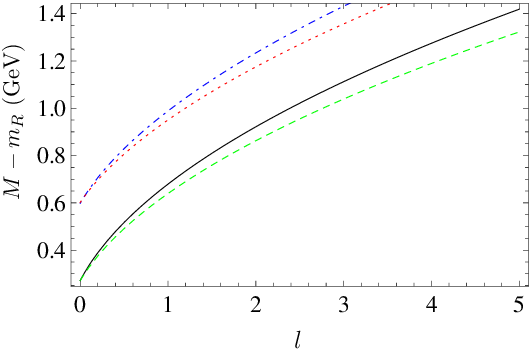}}
\caption{The radial and orbital {\rts} for the $(cu)$ and $(bu)$ diquarks, respectively. (a) Radial {\rts} for the $[cu]^{\bar{3}_c}_{1^1s_0}$ state and for the $[bu]^{\bar{3}_c}_{1^1s_0}$ state. (b) Radial {\rts} for the $\{cu\}^{\bar{3}_c}_{1^3s_1}$ state and for the $\{bu\}^{\bar{3}_c}_{1^3s_1}$ state. (c) Orbital {\rts} for the $[cu]^{\bar{3}_c}_{1^1s_0}$ state and for the $[bu]^{\bar{3}_c}_{1^1s_0}$ state. (d) Orbital {\rts} for the $\{cu\}^{\bar{3}_c}_{1^3s_1}$ state and for the $\{bu\}^{\bar{3}_c}_{1^3s_1}$ state. The black lines are the $(cu)$ {\rts} for Fit1, the green dashed lines are the $(bu)$ {\rts} for Fit1, the blue dot-dashed lines are the $(cu)$ {\rts} for Fit2 and the red dotted lines are the $(bu)$ {\rts} for Fit2. The data are listed in Tables \ref{tab:bcur} and \ref{tab:bcuo}.}\label{fig:bcu}
\end{figure*}

\begin{table*}[!phtb]
\caption{Comparison of theoretical predictions for the masses of the heavy-light diquarks (in {\gev}).}  \label{tab:cpmass}
\centering
\begin{tabular*}{0.8\textwidth}{@{\extracolsep{\fill}}ccccccccc@{}}
\hline\hline
 $j^P$ & Diquark  &Fit1 & Fit2  & FGS \cite{Faustov:2021hjs}
 & YCRS \cite{Yin:2021uom} & GTB \cite{Gutierrez-Guerrero:2021rsx}
 & G \cite{Giannuzzi:2019esi} & GAR \cite{Gutierrez-Guerrero:2021fuj} \\
\hline
$0^+$   &  $[cq]^{{\cltba}}_{1^1s_0}$
                  & 1.92  & 1.92   & 1.973    & 2.15  & 2.08  & 2.118 & 1.88 \\
        &  $[cs]^{{\cltba}}_{1^1s_0}$
                  & 2.04  & 2.03   & 2.091    & 2.26  & 2.17  & 2.237 & 2.0 \\
        &  $[bq]^{{\cltba}}_{1^1s_0}$
                  & 5.31  & 5.30   & 5.359    & 5.51  & 5.37  & 5.513 & 5.31 \\
        &  $[bs]^{{\cltba}}_{1^1s_0}$
                  & 5.42  & 5.41   & 5.462    & 5.60  & 5.46  & 5.619 & 5.40 \\
$0^-$   & $[cq]^{{\cltba}}_{1^3p_0}$
                   &2.32   & 2.30   &                & 2.35  & 2.37  &   \\
        & $[cs]^{{\cltba}}_{1^3p_0}$
                   &2.45   & 2.44   &                & 2.48  & 2.47  &   \\
        & $[bq]^{{\cltba}}_{1^3p_0}$
                   & 5.69  & 5.67   &                & 5.61  & 5.53  &   \\
        & $[bs]^{{\cltba}}_{1^3p_0}$
                   & 5.81  & 5.79   &                & 5.72  & 5.62  &   \\
%%%
$1^+$   & $\{cq\}^{{\cltba}}_{1^3s_1}$
                  & 2.00  & 2.00   & 2.036    & 2.24  & 2.16  & 2.168 & 2.03 \\
        & $\{cs\}^{{\cltba}}_{1^3s_1}$
                  & 2.12  & 2.12   & 2.158    & 2.34  & 2.25  & 2.276 &2.14 \\
        & $\{bq\}^{{\cltba}}_{1^3s_1}$
                  & 5.33  & 5.33   & 5.381    & 5.53  & 5.39  & 5.526 & 5.36 \\
        & $\{bs\}^{{\cltba}}_{1^3s_1}$
                  & 5.45  & 5.44   & 5.482    & 5.62  & 5.47  & 5.630 & 5.45 \\
$1^-$   & $\{cq\}^{{\cltba}}_{1^1p_1}$
                   & 2.38  & 2.36   &          & 2.45  & 2.45  &   \\
        & $\{cs\}^{{\cltba}}_{1^1p_1}$
                   & 2.53  & 2.51   &          & 2.56  & 2.54  &   \\
        & $\{bq\}^{{\cltba}}_{1^1p_1}$
                   & 5.69  & 5.67   &          & 5.67  & 5.59  &   \\
        & $\{bs\}^{{\cltba}}_{1^1p_1}$
                   & 5.81  & 5.80   &          & 5.77  & 5.67  &   \\
\hline\hline
\end{tabular*}
\end{table*}

Using Eq. (\ref{rtmeson}) with (\ref{massform}), (\ref{cxcons}) and (\ref{rtft}) (Fit1) and Eq. (\ref{mrtf}) with (\ref{mrfp}) (Fit2) to fit the radial {\rts} for the $1^1s_0$ ($1^3s_1$) charmed mesons and for the $1^1s_0$ ($1^3s_1$) bottom mesons, respectively, we obtain the parameters $c_{fn_r}$ and $c_{0n_r}$, see Table \ref{tab:fitparameters}.
%%%
The masses of mesons from PDG \cite{ParticleDataGroup:2022pth} and the theoretical predictions from \cite{Ebert:2009ua} are used to obtain $c_{fn_r}$. The experimental values \cite{ParticleDataGroup:2022pth} are used to obtain $c_{0n_r}$.
Substitute the values in Eq. (\ref{param}) and the obtained $c_{fn_r}$ and $c_{0x}$ into (\ref{rtmeson}), (\ref{massform}), (\ref{cxcons}), (\ref{rtft}), (\ref{mrtf}) and (\ref{mrfp}). Then the masses of diquark $(cu)$ and $(bu)$ can be calculated by Eqs. (\ref{rtmeson}) and (\ref{mrtf}), respectively, see Table \ref{tab:bcur}. The diquark masses calculated by Fit1 and Fit2 are approximately equal to each other.

Similar to the radial {\rt} case, the orbital {\rts} for the $1^1s_0$ ($1^3s_1$) charmed mesons and for the $1^1s_0$ ($1^3s_1$) bottom mesons are fitted by using Eq. (\ref{rtmeson}) with (\ref{massform}), (\ref{cxcons}) and (\ref{rtft}) (Fit1) and Eq. (\ref{mrtf}) with (\ref{mrfp}) (Fit2), repectively.
The experimental data from PDG \cite{ParticleDataGroup:2022pth} and the theoretical data from \cite{Ebert:2009ua} are used to obtain $c_{fl}$. The experimental values \cite{ParticleDataGroup:2022pth} are used to determine $c_{0l}$.
The fitted parameters are listed in Table \ref{tab:fitparameters}.
Using the fitted values, the diquark masses of the orbitally excited states can be calculated by using Eqs. (\ref{rtmeson}) and (\ref{mrtf}), respectively, see Table \ref{tab:bcuo}.
Here, we do not consider the mixtures of spin-triplet and spin-singlet states of the heavy-light diquarks. The $1^1l_l$ states are calculated by using the orbital {\rts} for the $1^1s_0$ state, see Table \ref{tab:bcuo}.

The calculated masses by using the {\rts} are in accordance with other theoretical predictions, see Table \ref{tab:cpmass}. In Table \ref{tab:cpmass}, Fit1 and Fit2 are obtained by using the orbital {\rts}, which are from Tables \ref{tab:bcuo} and \ref{tab:bcso}. By fitting the orbital {\rt} for $D^{\ast}_0(2300)$, we obtain fitted $c_{0l}$. Then the masses of $[cu]^{{\cltba}}_{1^3p_0}$ can be calculated.  By fitting the orbital {\rt} for the $1^3P_0$ state of the bottom meson, we have the value of $c_{0l}$. Then the masses of $[bu]^{{\cltba}}_{1^3p_0}$ can be calculated. The $1^3P_0$ state of the bottom meson has not been determined experimentally, therefore, its theoretical value from \cite{Ebert:2009ua} is used.

The radial and orbital {\rts} for the diquarks $(cu)$ and $(bu)$ are shown in Fig. \ref{fig:bcu}. In each figure, the upper two lines, fitted by using Eq. (\ref{mrtf}) with (\ref{mrfp}) (Fit2), are positioned above the lower two lines fitted by using Eq. (\ref{rtmeson}) with (\ref{massform}), (\ref{cxcons}) and (\ref{rtft}) (Fit1). This is because $m_R=m_1+m_2+C/2$ for Fit1 while $m_R=m_1+C/2$ for Fit2.
To compare the {\rts} for the $(cu)$ with those for the $(bu)$, they are plotted in the $(M-m_R,\,x)$ $(x=n_r,\,l)$ plane, see Fig. \ref{fig:bcu}. For both Fit1 and Fit2, the radial and orbital {\rts} for $(cu)$ and those for $(bu)$ almost overlap with each other irrespective of heavy quark flavors. This indicates an universal description of these heavy-light diquarks $(cu)$ and $(bu)$ which is similar to the universal description of the heavy-light mesons and baryons \cite{Chen:2023web,Jakhad:2023mni,Chen:2017fcs,Jia:2018vwl}.

\subsection{{\rts} for the $(cs)$ and $(bs)$ diquarks}

\begin{table}[!phtb]
\caption{Same as Table \ref{tab:bcur} except for the $(cs)$ and $(bs)$ diquarks. }  \label{tab:bcsr}
\centering
\begin{tabular*}{0.47\textwidth}{@{\extracolsep{\fill}}cccccc@{}}
\hline\hline
                      &   \multicolumn{2}{c}{$(cs)$} &   \multicolumn{2}{c}{$(bs)$}   \\
State ($n^{2s+1}l_j$) & Fit1 &  Fit2  & Fit1   &  Fit2 \\
\hline
$1^1s_0$           & 2.03    & 2.01    & 5.43    & 5.42      \\
$2^1s_0$           & 2.68    & 2.65    & 5.97    & 5.96      \\
$3^1s_0$           & 2.99    & 2.97    & 6.26    & 6.25      \\
$4^1s_0$           & 3.23    & 3.22    & 6.49    & 6.48      \\
$5^1s_0$           & 3.43    & 3.44    & 6.68    & 6.68      \\
%%%
$1^3s_1$           & 2.14    & 2.13    & 5.47    & 5.46    \\
$2^3s_1$           & 2.70    & 2.68    & 5.98    & 5.97    \\
$3^3s_1$           & 3.01    & 2.99    & 6.27    & 6.26    \\
$4^3s_1$           & 3.24    & 3.24    & 6.49    & 6.49    \\
$5^3s_1$           & 3.45    & 3.45    & 6.68    & 6.69    \\
\hline
\hline
\end{tabular*}
\end{table}

\begin{table}[!phtb]
\caption{Same as Table \ref{tab:bcuo} except for the $(cs)$ and $(bs)$ diquarks. }  \label{tab:bcso}
\centering
\begin{tabular*}{0.47\textwidth}{@{\extracolsep{\fill}}cccccc@{}}
\hline\hline
                      &   \multicolumn{2}{c}{$(cs)$} &   \multicolumn{2}{c}{$(bs)$}   \\
State ($n^{2s+1}l_j$) & Fit1 &  Fit2  & Fit1   &  Fit2 \\
\hline
$1^1s_0$         & 2.04    & 2.03  & 5.42  & 5.41  \\
$1^1p_1$         & 2.53    & 2.51  & 5.81  & 5.80  \\
$1^1d_2$         & 2.77    & 2.76  & 6.03  & 6.02  \\
$1^1f_3$         & 2.96    & 2.95  & 6.20  & 6.19  \\
$1^1g_4$         & 3.13    & 3.12  & 6.35  & 6.34  \\
$1^1h_5$         & 3.27    & 3.27  & 6.48  & 6.48  \\
%%%\hline
$1^3s_1$         & 2.12    & 2.12  & 5.45  & 5.44  \\
$1^3p_2$         & 2.55    & 2.53  & 5.82  & 5.81  \\
$1^3d_3$         & 2.79    & 2.78  & 6.04  & 6.03  \\
$1^3f_4$         & 2.98    & 2.97  & 6.21  & 6.20  \\
$1^3g_5$         & 3.14    & 3.14  & 6.35  & 6.35  \\
$1^3h_6$         & 3.28    & 3.29  & 6.48  & 6.48  \\
\hline\hline
\end{tabular*}
\end{table}

\begin{figure*}[!phtb]
\centering
\subfigure[]{\label{figcca}\includegraphics[scale=0.8]{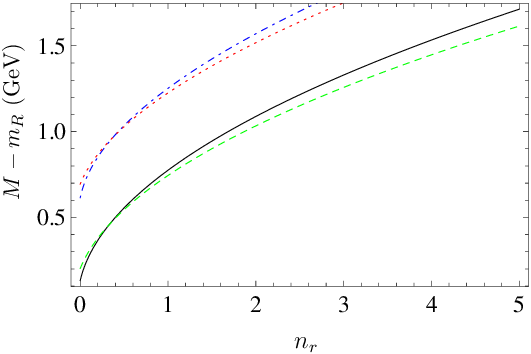}}
\subfigure[]{\label{figcca}\includegraphics[scale=0.8]{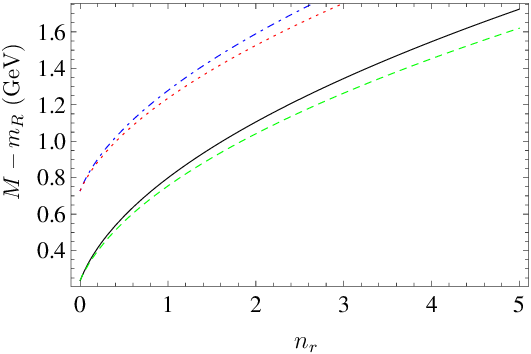}}
\subfigure[]{\label{figcca}\includegraphics[scale=0.8]{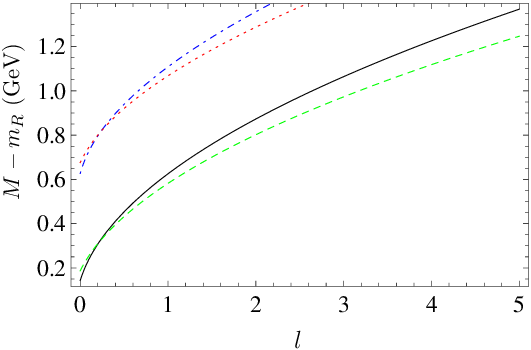}}
\subfigure[]{\label{figcca}\includegraphics[scale=0.8]{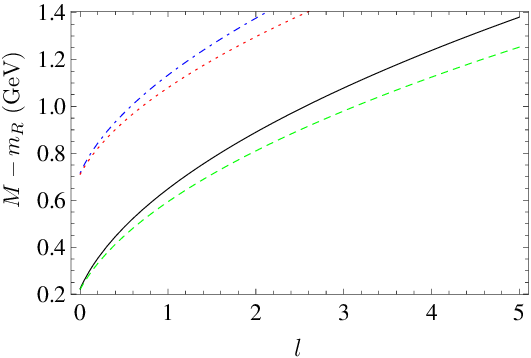}}
\caption{Same as Fig. \ref{fig:bcu} except for the diquarks $(cs)$ and $(bs)$.}\label{fig:bcs}
\end{figure*}

By using Eq. (\ref{rtmeson}) with (\ref{massform}), (\ref{cxcons}) and (\ref{rtft}) (Fit1) and Eq. (\ref{mrtf}) with (\ref{mrfp}) (Fit2) to fit the radial {\rts} for the $1^1s_0$ ($1^3s_1$) charmed-strange mesons and for the $1^1s_0$ ($1^3s_1$) bottom-strange mesons, respectively, the parameters $c_{fn_r}$ and $c_{0n_r}$ can be determined, see Table \ref{tab:fitparameters}.
%%%
The experimental data from PDG \cite{ParticleDataGroup:2022pth} and the theoretical data from \cite{Ebert:2009ua} are used to obtain $c_{fn_r}$. The experimental values \cite{ParticleDataGroup:2022pth} are used to obtain $c_{0n_r}$.
Substitute the values in Eq. (\ref{param}) and the obtained $c_{fn_r}$ and $c_{0x}$ into (\ref{rtmeson}), (\ref{massform}), (\ref{cxcons}), (\ref{rtft}), (\ref{mrtf}) and (\ref{mrfp}). Then the masses of diquark $(cs)$ and $(bs)$ can be calculated by Eqs. (\ref{rtmeson}) and (\ref{mrtf}), respectively. The obtained masses are listed in Tables \ref{tab:bcsr}. The diquark masses calculated by Fit1 and Fit2 are approximately equal to each other.

Similar to the radial {\rt} case, the orbital {\rts} for the $1^1s_0$ ($1^3s_1$) charmed-strange mesons and for the $1^1s_0$ ($1^3s_1$) bottom-strange mesons are fitted by using Eq. (\ref{rtmeson}) with (\ref{massform}), (\ref{cxcons}) and (\ref{rtft}) (Fit1) and Eq. (\ref{mrtf}) with (\ref{mrfp}) (Fit2), respectively, see Table \ref{tab:fitparameters}.
The experimental data from PDG \cite{ParticleDataGroup:2022pth} and the theoretical data from \cite{Ebert:2009ua} are used to obtain $c_{fl}$. The experimental values \cite{ParticleDataGroup:2022pth} are used to determine $c_{0l}$.
Using the fitted parameters, the diquark masses of the orbitally excited
states can be calculated by using Eqs. (\ref{rtmeson}) and (\ref{mrtf}), respectively, see Table \ref{tab:bcso}.

The calculated masses of the $(cs)$ and $(bs)$ diquarks by using the {\rts} are in accordance with other theoretical predictions, see Table \ref{tab:cpmass}. In Table \ref{tab:cpmass}, Fit1 and Fit2 are obtained by using the orbital {\rts}, which are from Tables \ref{tab:bcuo} and \ref{tab:bcso}.
By fitting the orbital {\rt} for $D^{\ast}_{s0}(2317)$, we have the fitted $c_{0l}$. Then the masses of $[cs]^{{\cltba}}_{1^3p_0}$ can be calculated.  By fitting the orbital {\rt} for the $1^3P_0$ state of the bottom-strange meson, we have the values of $c_{0l}$. Then the masses of $[bs]^{{\cltba}}_{1^3p_0}$ can be calculated. The $1^3P_0$ state of the bottom-strange meson has not been determined experimentally, therefore, its theoretical value from \cite{Ebert:2009ua} is used.

The radial and orbital {\rts} for the diquarks $(cs)$ and $(bs)$ are shown in Fig. \ref{fig:bcs}. In every graph in Fig. \ref{fig:bcs}, two lines by Fit2 lie above two lines by Fit1 because the light quark mass $m_2$ is not included in $m_R$ for Fit2.
For both Fit1 and Fit2, the radial and orbital {\rts} for $(cs)$ and those for $(bs)$ are close to each other irrespective of heavy quark flavors. This means the universal description of the charmed-strange and bottom-strange diquarks. We can conclude that the heavy-light diquarks $(cs)$ and $(bs)$ satisfy the universal descriptions same as other heavy-light systems, such as the heavy-light mesons, the heavy-light baryons composed of one heavy quark (diquark) and one light diquark (quark), and the heavy-light tetraquarks composed of one heavy diquark (anidiquark) and one light antidiquark (diquark) \cite{Chen:2023web,Jakhad:2023mni,Chen:2017fcs,Jia:2018vwl}.

\subsection{Discussions}
For the heavy-light mesons with $J=l$, there are mixing of the spin-triplet states and spin-singlet states \cite{Ebert:2009ua},
\begin{align}
|\psi_J\rangle&=|^1l_l\rangle cos\theta + |^3l_l\rangle sin\theta,\nonumber\\
|\psi'_J\rangle&=-|^1l_l\rangle sin\theta + |^3l_l\rangle cos\theta.
\end{align}
The mixing occurs due to the nondiagonal spin-orbit and tensor terms. Similar to the heavy-light mesons, there will exist the $^1l_l-^3l_l$ mixing for the heavy-light diquarks. In this work, we do not consider this kind of $^1l_l-^3l_l$ mixing. The masses of the $1^1l_l$ states of the heavy-light diquarks are estimated by using the orbital {\rts} for the $[qq']^{\cltba}_{1^1s_0}$ state, $qq'=cu,cs,bu,bs$, see Tables \ref{tab:bcuo} and \ref{tab:bcso}.

\begin{figure}[!phtb]
\centering
\includegraphics[scale=0.8]{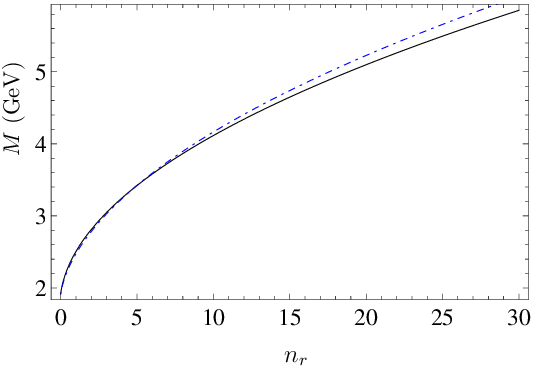}
\caption{The radial {\rts} for the $[cu]^{\bar{3}_c}_{1^1s_0}$ state fitted by Fit1 (the black line) and Fit2 (the blue dot-dashed line). The values of parameters are in \ref{subsec:pre} and \ref{subsec:cbu}.}\label{fig:fitc}
\end{figure}

The results obtained from Fit1 and Fit2 show good agreement when $x$ ($x=n_r,l$) is smaller than approximately 10. As $x$ increases, the values from Fit2 become greater than the values from Fit1. The differences between the values obtained from Fit1 and Fit2 increase with $x$. These situations vary with different {\rts}. To illustrate this, Fig. \ref{fig:fitc}, taken as an example, present the radial {\rts} for the $[cu]^{\bar{3}_c}_{1^1s_0}$ state fitted by Fit1 and Fit2.
It can be observed that Eq. (\ref{rtmeson}) with (\ref{massform}), (\ref{cxcons}) and (\ref{rtft}) (Fit1) and Eq. (\ref{mrtf}) with (\ref{mrfp}) (Fit2) are suitable for describing the heavy-light diquarks when $n_r<10$. In most of cases, there are limited experimental and theoretical data available for the very highly excited states with $n_r,l>10$, for example, for the heavy-light mesons or baryons. This implies that both {\rt} formulas can be applied to discussed the {\rts} for the heavy-light diquarks.

\section{Conclusions}\label{sec:conc}

We attempt to apply the Regge trajectory approach to investigate the heavy-light diquarks composed of one heavy quark and one light quark. The spectra of the heavy-light diquarks $(cu)$, $(cs)$, $(bu)$ and $(bs)$ are calculated by using the {\rt} approach, and these results are in agreement with other theoretical results. This demonstrates the appropriateness of the {\rt} method for studying the heavy-light diquarks. The diquark Regge trajectory provides a new and simple approach for estimating the spectra of the heavy-light diquarks. Additionally, we expected that the {\rt} approach can also be applied to the light diquarks composed of two light quarks.

The {\rt} relation for the doubly heavy diquarks has the same form as the relation for the doubly heavy mesons. However, unlike the case of the doubly heavy diquark, the usual {\rt} formula for the heavy-light mesons cannot be directly applied to the heavy-light diquarks. We find that considering the light quark mass and the parameter $C$ is necessary to obtain agreeable results. Two modified formulas present unified descriptions of the heavy-light mesons and the heavy-light diquarks.

We present a method for determining the parameters in the diquark {\rts}. By employing  (\ref{rtmeson}) with (\ref{massform}), (\ref{cxcons}) and (\ref{rtft}) or Eq. (\ref{mrtf}) with (\ref{mrfp}) to fit the heavy-light mesons, we can obtain values for the universal parameters. By fitting a chosen meson {\rt}, $c_{0x}$ is calculated. Once all parameters are computed, the {\rt} for the heavy-light diquarks is determined, and their spectra can be estimated.

It is illustrated that the heavy-light diquarks exhibit an universal description, irrespective of heavy quark flavors, similar to other heavy-light systems such as the heavy-light mesons, the heavy-light baryons and the heavy-light tetraquarks.

\section*{Acknowledgments}
We are very grateful to the anonymous referees for the
valuable comments and suggestions.

\end{document}